\documentclass{jpp}
\usepackage{amsmath}
\usepackage{textcomp,gensymb}
\usepackage[pdftex]{graphicx}
\usepackage[pdftex]{color}
\usepackage{natbib}
\usepackage{units}
\usepackage{amssymb}
\usepackage{xspace}

\newcommand{\about}{\ensuremath{\mathbin{\sim}}}
\newcommand{\vect}[1]{\ensuremath{\boldsymbol{#1}}}
\newcommand{\del}{\boldsymbol{\nabla}}
\newcommand{\cross}{\boldsymbol{\times}}
\newcommand{\ptl}{\partial}
\newcommand{\e}[1]{\times 10^{#1}}
\newcommand{\vorpal}{\textsc{vorpal}}

\newcommand{\vorpalpd}{\textsc{vorpal-pd}\xspace}
\newcommand{\calder}{\textsc{calder-circ}\xspace}

\title[Computationally efficient methods for modelling LWFA in the bubble regime]{Computationally efficient methods for modelling laser wakefield acceleration in the blowout regime}

\author[B. M. Cowan \emph{et al.}]%
{B.\ns M.\ns C\ls O\ls W\ls A\ls N$^1$%
	\thanks{Email address for correspondence: benc@txcorp.com},\ns
S.\ns Y.\ns K\ls A\ls L\ls M\ls Y\ls K\ls O\ls V$^2$%
	\thanks{Email address for correspondence: skalmykov2@unl.edu},\ns
A.\ns B\ls E\ls C\ls K$^3$\break
X.\ns D\ls A\ls V\ls O\ls I\ls N\ls E$^4$,\ns
K.\ns B\ls U\ls N\ls K\ls E\ls R\ls S$^2$,\ns
A.\ns F.\ns L\ls I\ls F\ls S\ls C\ls H\ls I\ls T\ls Z$^5$,\break
E.\ns L\ls E\ls F\ls E\ls B\ls V\ls R\ls E$^4$,\ns
D.\ns L.\ns B\ls R\ls U\ls H\ls W\ls I\ls L\ls E\ls R$^1$,\break
B.\ns A.\ns S\ls H\ls A\ls D\ls W\ls I\ls C\ls K$^2$,\ns
\and D.\ns P.\ns U\ls M\ls S\ls T\ls A\ls D\ls T\ls E\ls R$^2$}

\affiliation{$^1$Tech-X Corporation, 5621 Arapahoe Ave. Ste. A, Boulder, CO 80303, USA\\[\affilskip]
$^2$Department of Physics and Astronomy, University of Nebraska -- Lincoln, NE 68588-0299, USA\\[\affilskip]
$^3$Centrum voor Plasma-Astrofysica, Departement Wiskunde, Katholieke Universiteit Leuven, Celestijnenlaan 200B, B-3001 Leuven, Belgium\\[\affilskip]
$^4$CEA, DAM, DIF, Arpajon F-91297, France\\[\affilskip]
$^5$Laboratoire d'Optique Appliqu\'ee, ENSTA, Ecole Polytechnique, CNRS, 91761 Palaiseau, France}

\pubyear{2012}
\volume{???}
\pagerange{???--???}
\date{?; revised ?; accepted ?. - To be entered by editorial office}

\begin{document}

\maketitle

\begin{abstract}
Electron self-injection and acceleration until dephasing in the blowout regime is studied for a set of initial conditions typical of recent experiments with 100 terawatt-class lasers.  Two different
approaches to computationally efficient, fully explicit, three-dimensional particle-in-cell modelling are examined.  First, the Cartesian code \vorpal\ \citep{Nieter:VORPAL} using a perfect-dispersion
electromagnetic solver precisely describes the laser pulse and bubble dynamics, taking advantage of coarser resolution in the propagation direction, with a proportionally larger time step.  Using
third-order splines for macroparticles helps suppress the sampling noise while keeping the usage of computational resources modest.  The second way to reduce the simulation load is using
reduced-geometry codes.  In our case, the quasi-cylindrical code \calder \citep{Lifschitz:CalderCirc} uses decomposition of fields and currents into a set of poloidal modes, while the macroparticles
move in the Cartesian 3D space.  Cylindrically symmetry of the interaction allow using just two modes, reducing the computational load to roughly that of a planar Cartesian simulation while
preserving the 3D nature of the interaction.  This significant economy of resources allows using fine resolution in the direction of propagation and a small time step, making numerical dispersion
vanishingly small, together with a large number of particles per cell, enabling good particle statistics.  Quantitative agreement of the two simulations indicates that they are free of numerical
artefacts.  Both approaches thus retrieve physically correct evolution of the plasma bubble, recovering the intrinsic connection of electron self-injection to the nonlinear optical evolution of the
driver.
\end{abstract}

\section{Introduction}

Relativistic Langmuir waves driven by short, intense laser pulses in rarefied plasmas maintain accelerating gradients several orders of magnitude higher than those accessible in conventional metallic
structures \citep{Tajima:LaserElectronAcc,Gorbunov:Early3DWake,Esarey:LPAPhysicsReview}.  The technical simplicity and compactness of these laser-plasma accelerators (LPAs) is attractive for a broad
range of applications, such as nuclear activation and on-site isotope production \citep{Leemans:NuclearActivation,Reed:NuclearActivation}, long-distance probing of defects in shielded structures
\citep{Ramanathan:XrayRadiography}, and testing radiation resistivity of electronic components \citep{Hidding:CosmicRays}.  Realisation of compact, inexpensive, bright x- and gamma-ray sources using
electron beams from LPAs \citep{Rousse:Production,Rousse:Scaling,Phuoc:LaserSynchrotron,Kneip:BrightSynchrotron,Cipiccia:ForcedSynchrotron} holds the promise to enable a much wider user community than
can be served by existing large-scale facilities.  These applications are not especially demanding as regards electron beam quality, and in fact sometimes draw benefits from poor beam collimation and
a broad energy spectrum \citep{Hidding:CosmicRays}.  However, there are also important applications with much tighter beam requirements.  Such applications include generating coherent x-rays using an
external magnetic undulator \citep{Gruener:DesignConsiderations,Schlenvoigt:CompactSynchrotron,Schlenvoigt:SynchrotronRadiation,Fuchs:SoftXRay}, producing x-rays for the phase contrast imaging
\citep{Fourmaux:PhaseContrastALLS,Kneip:PhaseContrast}, and high-brightness, quasi-monochromatic gamma-ray Compton sources \citep{Leemans:RadiationFromLPA,Hartemann:ComptonScattering}; these require
electron beams with a multi-kA current, low phase space volume, and energy in the few-gigaelectronvolt (GeV) range.

Achieving this high level of accelerator performance is a major near-term goal of the LPA community.  Modern laser systems capable of concentrating up to 10 Joules of energy in a sub-50 femtosecond
pulse \citep{Yanovsky:HERCULES,Froula:GeV,Kneip:GeV,Fourmaux:PhaseContrastALLS} make it possible to achieve the so-called blowout (or ``bubble'') regime, which is desirable due to its technical
simplicity and scalability \citep{Gordienko:Scalings,Lu:Generating}.  In this regime, motion of the electrons in the focus of the laser pulse is highly relativistic.  The laser ponderomotive force
expels plasma electrons from the region of the pulse, while the fully stripped ions remain essentially immobile, creating a column of positive charge in the laser wake.  The charge separation force
attracts bulk plasma electrons to the axis, creating a closed bubble devoid of electrons.  This co-propagating electron density bubble
\citep{Rosenzweig:Blowout,Mora:SelfChanneling,Pukhov:LWFABroken,Gordienko:Scalings,Lu:BubbleTheory} guides the laser pulse over many Rayleigh lengths \citep{Mora:SelfChanneling,Lu:Generating}.  The
bubble readily traps initially quiescent background electrons, accelerating them to hundreds of megaelectronvolts (MeV) over a few mm, creating a collimated electron bunch
\citep{Pukhov:LWFABroken,Kalmykov:SelfInjectionPhysPlasmas}.  It is in this regime that the first quasi-monoenergetic electrons were produced from laser plasmas in the laboratory
\citep{Geddes:HighQuality,Mangles:Monoenergetic,Faure:Monoenergetic}, and the GeV energy range was approached
\citep{Leemans:GeV,Karsch:GeVScale,Hafz:StableGeneration,Froula:GeV,Kneip:GeV,Clayton:GeVIonizationInducedInjection,
Lu:GeVChannelIonization,Liu:GeVIonizationInducedInjection,Pollock:GeVIonizationInducedInjection}.

Multi-dimensional particle-in-cell (PIC) simulations have played a key role in understanding the physics of the fully kinetic, strongly relativistic blowout regime.  The PIC method
\citep{Hockney:Particles,Birdsall:PlasmaSimulation} self-consistently models both electromagnetic fields and charged particles, representing field quantities on a grid and particles in a continuous
phase space.  Given sufficient computing power, electromagnetic PIC codes can simulate the plasma electrons (and ions, if necessary), the laser pulse driving the plasma wake, and the dynamics of
electrons injected into the accelerating potential.  In particular, two- and three-dimensional PIC simulations have been essential in understanding the dynamical nature of the electron self-injection
process \citep{Xu:ExpandingBubble,Oguchi:OscillatingBubble,Wu:InjectionFromSheath,Zhidkov:OscillatingBubble,
Kalmykov:ElectronSelfInjection,Kalmykov:NumericalModelling,Kalmykov:SelfInjectionPhysPlasmas,Kalmykov:InTechBook,Kalmykov:PlasmaLens}.  However, to capture precisely the correlation between driver
dynamics, electron self-injection, and GeV-scale acceleration in the bubble regime, a simulation must meet a number of challenging requirements.

Optimisation of a GeV-scale LPA performance, even with the use of massively parallel computation, is a challenging task especially because of the necessary cm-scale laser-plasma interaction length.
The laser energy is used most effectively if electrons are accelerated until they outrun the bubble and exit the accelerating phase, at which point they will have gained the maximum possible energy in
an LPA stage,
\begin{equation}
E_d\approx \unit[2.7\gamma_g^{4/3}P_{\text{TW}}^{1/3}]{MeV}.
\label{eq:Ed}
\end{equation}
Acceleration to this \textit{dephasing limit} occurs over the distance \citep{Lu:Generating}
\begin{equation}
L_d\approx 0.6\lambda_0\gamma_g^{8/3}P_{\text{TW}}^{1/6}.
\label{eq:Ld}
\end{equation}
Here, $P_{\text{TW}}$ is the laser power in terawatts ($\unit[1]{TW} = \unit[10^{12}]{W}$), $\gamma_g = \omega_0/\omega_{\text{pe}}\gg 1$ is the Lorentz factor associated with the linear group
velocity of the pulse in plasma, $\omega_0$ is the laser frequency, $\lambda_0 = 2\pi c/\omega_0$ is the laser wavelength, $\omega_{\text{pe}} = (n_0e^2/m_e\epsilon_0)^{1/2}$ is the electron Langmuir
frequency, $m_e$ is the electron rest mass, $n_0$ is the background electron density, $e$ is the electron charge, and $\epsilon_0$ is the permittivity of free space.  The scalings \eqref{eq:Ed} and
\eqref{eq:Ld} imply that the pulse remains self-guided, namely, that it remains longer than $c\, \omega_{\text{pe}}^{-1}$ \citep{Sprangle:NonlinearInteraction,Gorbunov:SuppressionOfSelffocusing}, and
that its power exceeds the critical power for relativistic self-focusing, $\unit[P_{\text{cr}}=16.2\gamma_g^2]{GW}$ \citep{Sun:SelfFocusing}.  Increasing the electron energy therefore requires
reduction of the electron plasma density, increasing both the bubble velocity and size,
\begin{equation}
L_{\text{acc}}\approx 0.9\lambda_0\gamma_g^{2/3}P_{\text{TW}}^{1/6},
\label{eq:Rb}
\end{equation}
where $L_{\text{acc}}$ is the length of the accelerating phase of the wakefield (roughly equal to the bubble radius).  Electron dephasing scales as $L_d\sim n_0^{-4/3}$ and thus the final energy gain
scales as $E_d\sim n_0^{-2/3}$.  For instance, reaching \unit[1]{GeV} energy with a \unit[200]{TW} pulse and a wavelength of $\lambda_0 = \unit[0.8]{\micro m}$ may be achieved in a \unit[0.47]{cm}
length plasma with density $n_0 = \unit[3.5\times10^{18}]{cm^{-3}}$, and doubling that energy would require nearly four times the plasma length and three times lower density, also increasing the
bubble size by \about 40\%.  Simulations of LPA commonly use a moving-widow, where the simulation box propagates with the speed of light colinearly with the laser pulse.  This optimisation
notwithstanding, even the experiments with currently operating \unit[100]{TW} systems bring forth the task of modelling the pulse propagation in cm-length plasmas, with the size of 3D simulation box
on the order of hundred(s) of microns longitudinally and transversely.

The greatest challenge arises from the great disparity of physical scales between the laser wavelength and plasma length, which is the hallmark of high-energy laser-plasma acceleration.  The need to resolve
the laser wavelength, $\lambda_0\sim\unit[1]{\micro m}$, fixes the grid resolution, and, due to stability conditions \citep{Courant:1967aa}, also limits the time step to a small fraction of
$\omega_0^{-1}$.  Furthermore, the strong localisation of the injection process imposes even stricter limit on grid resolution; the vast majority of injection candidates are concentrated in the inner
lining of the bubble (the sheath), and penetrate into the bubble near its rear, where the sheath is longitudinally compressed to a few tens of nanometres
\citep{Wu:InjectionFromSheath,Kalmykov:SelfInjectionPhysPlasmas}.  Resolving this structure, together with ensuring sufficient particle statistics in the sheath, is necessary to avoid excessive
sampling noise and eliminate unphysical effects.  In this situation, extending the plasma length to centimeters and increasing the size of the simulation window to hundreds of microns, while at the
same time maintaining sufficient macroparticle statistics, would require solving Maxwell's equations on meshes amounting to billions of grid points, and advancing 1--10 billion macroparticles over
millions of time steps.  Performing such simulations with standard electromagnetic solvers and particle movers requires a national-scale supercomputing facility.  As a result, an attempt to reproduce
the long time scale evolution of the laser and the bubble together with fine details of the electron self-injection dynamics is usually a compromise between affordable simulation load and unavoidable
coarseness of the results.  However, the high precision of modern LPA experiments and high beam quality requirements of the applications are rather unforgiving to these compromises and do not tolerate
numerical artefacts \citep{Cormier-Michel:Unphysical}.

These considerations make it clear that PIC algorithms must be modified in order to reduce the required computational resources without compromising precision.  One of the main directions is
development of electromagnetic solvers that minimize numerical error while using the lowest possible grid resolution.  One particular limitation of PIC that requires high longitudinal resolution is
that of numerical dispersion.  In PIC, electromagnetic fields are typically updated using the finite-difference time-domain (FDTD) method on a staggered Yee grid \citep{Yee:Grid,Taflove:FDTD3}.  This
method is second-order accurate, and since it is explicit and local, it parallelizes efficiently enabling large-scale simulations.  However, it is known that this algorithm experiences numerical
dispersion error for waves propagating along the axis, which leads to errors in the group velocity of the laser pulse.  This artificial slowdown of the driver and the bubble leads to incorrect
dephasing of accelerated electrons and also permits synchronisation of the sheath electrons with the bubble, leading to their unphysical injection.  Mitigating this effect by using higher resolution
increases the computation time quadratically.  Because of the deleterious effects of numerical dispersion in FDTD schemes, efforts have been made to develop \emph{perfect dispersion} algorithms, which
exhibit no numerical dispersion for waves propagating along a grid axis.  For accelerator applications, several modifications to FDTD have been described that correct for numerical dispersion using
implicit methods \citep{Zagorodnov:RelSource,Zagorodnov:TETM}.  Because LPA simulations tend to be quite large-scale (using thousands of processor cores), an explicit algorithm is desirable for
reasons of computational efficiency.  Such an algorithm has been described in 2D \citep{Pukhov:VLPL} and in 3D for cubic cells \citep{Karkkainen:LowDispersion}.  These algorithms have also been
explored for LPA as a means of reducing noise in boosted-frame simulations \citep{Vay:InstabilityMitigation}.

In this paper, we use two complementary simulation codes (with different numerical approaches and physics content) to explore physical phenomena involved in self-injection and acceleration of
electrons until dephasing under typical conditions of recent experiments with \unit[100]{TW}-class lasers.  We use a newly-developed perfect-dispersion algorithm \citep{benc:PerfDispInPrep}
implemented in the fully explicit 3D Cartesian \vorpal\ simulation framework \citep{Nieter:VORPAL}, subsequently referred to as \vorpalpd.  The algorithm, briefly described in
Sec.~\ref{sec:perfectDispersion}, eliminates numerical dispersion in the direction of pulse propagation.  Thus, even with a relatively large longitudinal grid spacing (\about 15 grid points per
$\lambda_0$), the correct group velocity of a broad-bandwidth laser pulse is obtained.

The other code used here, \calder, uses cylindrical geometry.  This code uses poloidal mode decomposition of fields and currents defined on a radial grid, while macro\-particles retain their full 3D
dynamics in Cartesian coordinates \citep{Lifschitz:CalderCirc}.  Well-preserved cylindrical symmetry of the laser-plasma interaction enables using just a few lower-order modes.  Neglecting
higher-order, non-axisymmetric contributions to the wakefields and currents makes it possible to approach the performance of a 2D code.  \calder thus allows for fast, extra-high resolution runs with
excellent macroparticle statistics \citep{Kalmykov:NumericalModelling,Kalmykov:SelfInjectionPhysPlasmas}.

The paper is organized as follows.  In section \ref{sec:perfectDispersion} we outline the main features of the recently implemented perfect dispersion algorithm in the \vorpalpd code.  Section
\ref{sec:comprehensiveBenchmarking} is dedicated to the benchmarking of \vorpalpd against \calder.  Sec.~\ref{sec:conclusions} summarizes the results and indicates the directions of future work.

\section{The perfect dispersion method\label{sec:perfectDispersion}}

In this section we give a brief overview of the perfect dispersion method we use; a more complete description together with detailed benchmarks will be presented in \citep{benc:PerfDispInPrep}.  Our
method
is based on that in \citep{Pukhov:VLPL,Karkkainen:LowDispersion}, in which the FDTD algorithm is modified by smoothing the fields in the curl operator in one of Maxwell's equations.  We choose to
smooth the
electric fields for the magnetic field update; our update equations are then
\begin{equation}
	D_t\vect{B} = -\del'\cross\vect{E}, \quad D_t\vect{E} = c^2\del\cross\vect{B} - \frac{\vect{J}}{\epsilon_0},
\label{eq:perfDispMaxwell}
\end{equation}
where $\vect{J}$ is the electric current deposited from particle motion.  Here $D_t$ is the finite difference time derivative, $\del\cross$ is the standard finite difference curl operator, and
$\del'\cross$ is the modified curl operator.  Our modification to the curl operator involves applying smoothing transverse to the coordinate axis along which the derivative is taken.  For instance,
when computing $\ptl E_y/\ptl x$, $E_y$ is smoothed in the $y$ and $z$ directions.  This is equivalent to applying a smoothing operator before the numerical derivative operator.  The electric field
is smoothed only for the update of the magnetic field; the smoothed fields are not stored for the next time step.

The smoothed curl operator $\del'\cross$ is formed by modifying the finite difference operation.  If $D_i$ is the numerical derivative operator in the $i$-th direction, then for the modified curl we
use
$D_iS_i$ in place of $D_i$, where $S_i$ is the smoothing operator for the derivative.  The smoothing operator $S_x$ is defined by the stencil in the $y$ and $z$ directions
\begin{equation}
\begin{bmatrix}
\gamma_{yz} & \beta_z & \gamma_{yz} \\
\beta_y & \alpha_x & \beta_y \\
\gamma_{yz} & \beta_z & \gamma_{yz}
\end{bmatrix},
\end{equation}
and similar relations hold for cyclic permutations of the coordinate indices.  The coefficients $\alpha_i$, $\beta_i$, and $\gamma_{ij}$ are chosen to guarantee that waves propagating along the $x$
axis
(the laser propagation direction in our simulations) in vacuum experience no numerical dispersion, as described in \citep{benc:PerfDispInPrep}.  The only constraint is that the longitudinal grid
spacing
$\Delta x$ must satisfy $\Delta x \le \Delta y, \Delta z$ for the transverse grid spacings $\Delta y$ and $\Delta z$.

\section{Benchmarking\label{sec:comprehensiveBenchmarking}}

While a technological path to high-quality GeV beams exists, experimental progress is impeded by an incomplete understanding of the intrinsic relation between electron self-injection and nonlinear
optical
evolution of the driver, and hence by the lack of suitable criteria for selection of the optimal regimes that produce beams with the smallest possible phase-space volume. Control and optimisation of
the
fully kinetic, intrinsically 3D process of electron self-injection is a daunting task. It involves a systematic study of the links among the dynamics of self-injection and the nonlinear optical
processes
involving the laser pulse and the bubble.

Due to the extended acceleration length, the interaction of the laser pulse with the plasma is rich in nonlinear phenomena.  Even a Gaussian beam which is perfectly matched to the electron density
gradient in which it propagates is not immune to nonlinear optical processes.  Oscillations of the pulse spot-size due to non-linear refraction
\citep{Oguchi:OscillatingBubble,Zhidkov:OscillatingBubble,Kalmykov:NumericalModelling}, self-phase modulation leading to the formation of a relativistically intense optical piston
\citep{Tsung:PulseCompression,Lontano:Dynamics,Faure:PulseCompression, Pai:PulseCompression,Vieira:PulseCompression,Kalmykov:SelfInjectionPhysPlasmas,Kalmykov:InTechBook}, and relativistic
filamentation \citep{Andreev:Filamentation,Thomas:EffectFocusing,Thomas:Filamentation} are processes which result in pulse deformations.  Electron self-injection appears to be extremely sensitive to
such changes in pulse shape, which lead to contamination of the electron beam with polychromatic, poorly collimated background \citep{Kalmykov:InTechBook}.  Such contamination is readily seen even in
simulations with idealised initial conditions \citep{Kneip:GeV,Froula:GeV,Martins:Boost:NaturePhys,Kalmykov:NumericalModelling,Kalmykov:SelfInjectionPhysPlasmas}.  The complicated modal structure of
the incident pulse further aggravates the situation, leading to continuous off-axis injection, collective betatron oscillations \citep{Glinec:BetatronOffAxis,Mangles:Coma,Cummings:Aberrations}, and
electron beam steering \citep{Popp:AllOpticalSteering}.  In practice, these phenomena currently preclude operation reliable enough to enable high-precision user experiments; reported islands of stability for
self-injection in laser and plasma parameter space remain relatively narrow \citep{Karsch:GeVScale,Mangles:Stability,Thomas:EffectFocusing,Hafz:StableGeneration, Maksimchuk:FinalFOCUSReport,
Wiggins:Monoenergetic}.  Numerical codes used in predictive modelling of LPAs must be able to reproduce these phenomena with high precision in order not to confuse the instability of acceleration
caused by physical processes with unphysical artefacts caused by intrinsic deficiencies of numerical algorithms, such as numerical dispersion, high sampling noise, and grid heating.

\subsection{Simulation parameters\label{sec:parameters}}

The simulations presented here extend the earlier case study by \citep{Kalmykov:SelfInjectionPhysPlasmas} and use the same set of initial conditions.  A transform-limited Gaussian laser pulse with
full width at half-maximum (FWHM) in intensity $\tau_L = \unit[30]{fs}$, wavelength $\lambda_0 = \unit[0.805]{\micro m}$, and \unit[70]{TW} power is focussed at the plasma border ($x = 0$) into a spot
size $r_0 = \unit[13.6]{\micro m}$, and propagates in the positive $x$ direction.  The laser pulse is polarised in the $y$ direction.  The peak intensity at the focus is $\unit[2.3\e{19}]{W/cm^2}$,
giving a normalised vector potential of $a_0 = 3.27$.  The plasma density has a \unit[0.5]{mm} linear entrance ramp followed by a \unit[2]{mm} plateau and a \unit[0.5]{mm} linear exit ramp.  The
density in the plateau region, $n_0 = \unit[6.5\e{18}]{cm^{-3}}$, corresponds to $\gamma_g\approx P/P_{\text{cr}}\approx 16.3$ and dephasing length $L_d\approx\unit[1.7]{mm}$.

The simulations carried out with \vorpalpd use grid spacings of $\Delta x = 0.06\lambda_0 = \unit[48.3]{nm}$ longitudinally and $\Delta y = \Delta z = 0.5\lambda = \unit[403]{nm}$ transversely, with
four
macroparticles per cell.  Use of third-order splines for the macroparticle shapes reduces the sampling noise, mitigating the adverse effect of the coarse grid.  The domain in the \vorpalpd simulation
is
$\unit[72]{\micro m}$ long and $\unit[91]{\micro m}$ wide, and is surrounded transversely by a 16-layer perfectly-matched layer absorbing boundary.  The code is fully parallelised, and was run using
6\,144
cores on the Hopper supercomputer at the National Energy Research Scientific Computing Center (NERSC).  Completion of a typical run took $\about 3\e{5}$ CPU hours.

The \calder simulation uses 45 macroparticles per cylindrical cell, formed by the revolution of the grid cell around the propagation axis.  The longitudinal grid spacing is $\Delta x =
0.125c/\omega_0\approx \unit[16]{nm}$.  The aspect ratio $\Delta r/\Delta x = 15.6$ (where $r=\sqrt{y^2 + z^2}$), and the time step $\Delta t = 0.1244\omega_0^{-1}$.  With these grid parameters,
numerical
dispersion is negligible, and sampling noise is significantly reduced.  This high resolution simulation does not indicate any new physical effects compared to the \vorpalpd simulation, and does not
exhibit
significant differences in the quantitative results.  Well preserved cylindrical symmetry during the interaction (confirmed in the \vorpalpd simulation) enables us to approximate fields and currents
using
just the two lowest-order poloidal modes, thus reducing the 3D problem to an essentially 2D one.  These results confirm the earlier established fact \citep{Lifschitz:CalderCirc} that, in the case of a
linearly polarised laser, higher-order modes contribute only weakly to the electric field.  Comparison with the results of the \vorpalpd runs shows that our restriction to only two modes is
sufficiently
precise to reproduce all relevant physical effects, and to simulate the propagation through a \unit[3]{mm} plasma in 2\,625 CPU hours on 250 cores.

\subsection{Formation of quasi-monoenergetic bunches and physical origin of dark current\label{sec:actualBenchmarking}}

Upon entering the plasma, the strongly overcritical pulse rapidly self-focuses, reaching its highest intensity at $x\approx\unit[0.8]{mm}$, soon after entering the density plateau.  Full blowout is
maintained over the entire propagation distance.  In both simulations, electrons are accelerated until dephasing in two distinct stages, each characterised by completely different laser pulse
dynamics.  \emph{Transverse} evolution of the laser pulse is the hallmark of Stage I. The pulse spot size oscillates, first causing expansion and then contraction of the bubble.  The bubble expansion
produces self-injection of electrons from the sheath; stabilisation and contraction of the bubble extinguish injection, limiting the beam charge to a fraction of a nC. Phase space rotation creates a
well-collimated quasi-monoenergetic bunch long before dephasing.  Further acceleration (Stage II) is dominated by \emph{longitudinal (temporal)} self-compression of the pulse, leading to gradual
elongation of the bubble and continuous injection, producing a polychromatic, poorly collimated energy tail with a few nC charge.  This two-stage evolution has been noticed in earlier simulations
\citep{Froula:GeV,Kneip:GeV}, and explained in detail in \citep{Kalmykov:SelfInjectionPhysPlasmas}.

\begin{figure}
\begin{center}
\includegraphics{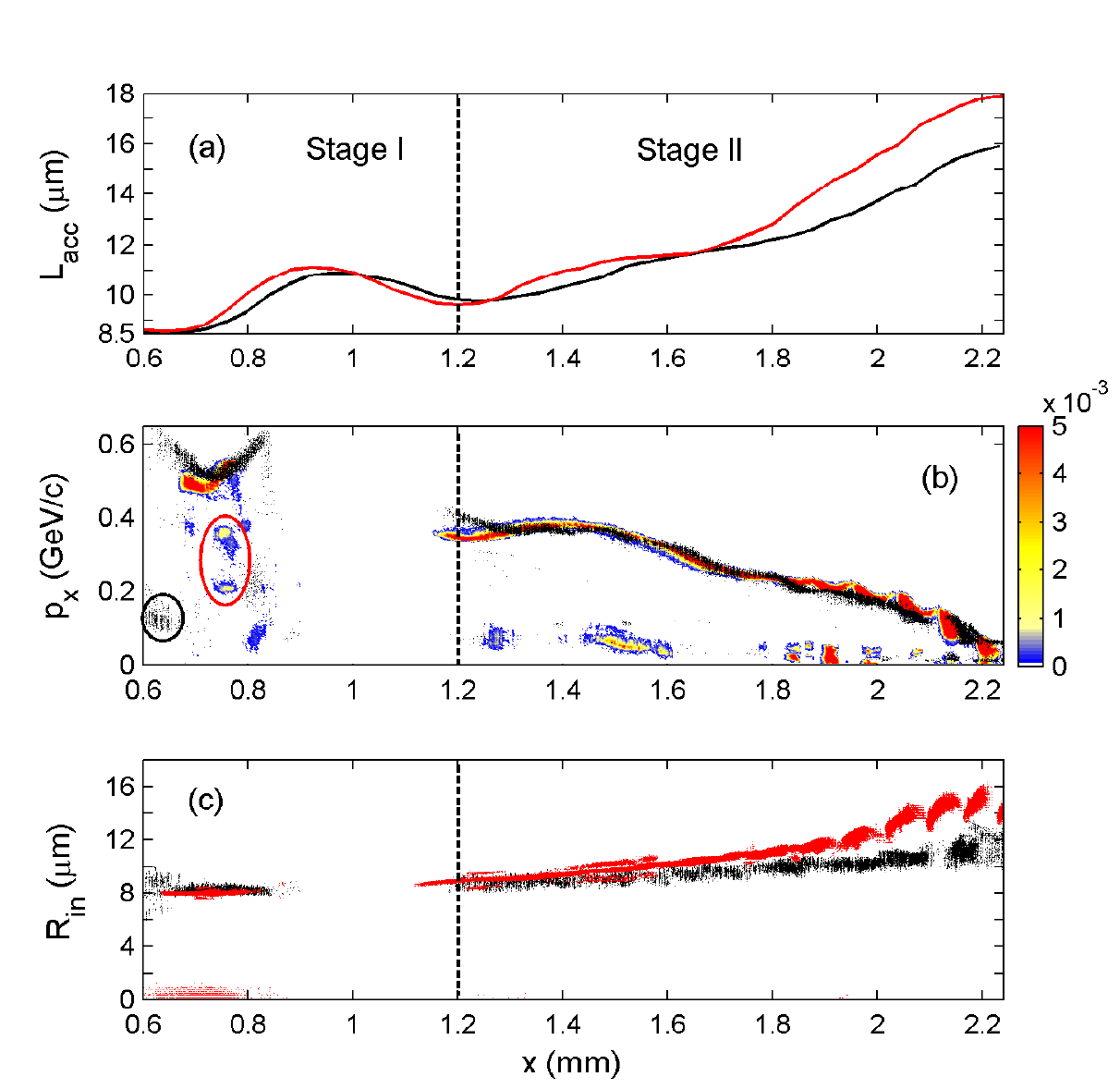} 
\caption{(a) Length of the accelerating phase vs.\ propagation distance in \calder (black) and \vorpalpd simulations (red/grey).  Expansion and contraction of the bubble due to nonlinear focusing of
the driver (Stage I) is followed
by continuous expansion caused by pulse self-compression (Stage II).  (b) Longitudinal momentum of electrons reaching the dephasing point, $x_{\text{deph}}\approx\unit[2.4]{mm}$, vs.\ their initial
longitudinal positions.  Black dots are the \calder macroparticles; the colourmap represents the normalised number density of \vorpalpd macroparticles.  Electrons are injected only during periods of
bubble
expansion.  A quasi-monoenergetic bunch forms during Stage I and maintains its low energy spread until dephasing, indicated by the group of early-injected particles with $E\approx\unit[500]{MeV}$.
Groups
of electrons encompassed by the ellipses were injected into the second and third buckets, to be further captured and accelerated by the expanding first bucket.  Continuous injection during Stage II
creates
a polychromatic energy tail.  (c) Collection volume: initial radial offsets of electrons reaching dephasing limit $R_{\mathrm{in}}=\sqrt{y_{\mathrm{in}}^2+z_{\mathrm{in}}^2}$, vs.\ their initial
longitudinal positions $x_{\mathrm{in}}$.  Black (red/grey) dots are \calder (\vorpalpd) macroparticles.  This collection volume indicates that the vast majority of electrons are collected from a
hollow
conical cylinder with a radius slightly smaller than the local bubble size.}
\label{fig:correlation}
\end{center}
\end{figure}

The correlation between the plasma bubble evolution and the self-injection process is quantified in figure~\ref{fig:correlation}.  Panel (a) shows the length of the accelerating phase on axis, viz.\
the length of the region inside the bubble where the longitudinal electric field is negative.  Panel (b) shows the longitudinal ``collection phase space'', viz.\ momenta of macroparticles reaching the
dephasing point, $p_{x}(x = x_{\text{deph}})$, vs.\ their initial position in plasma.  Panel (c) shows the collection volume: the initial positions of electrons reaching the dephasing point.
Comparison of these three panels shows that \emph{electrons are injected only during the periods of bubble expansion}.

During Stage I, radial oscillation of the laser pulse tail inside the bubble causes alternating expansion and contraction of the first bucket, clearly seen in the progression from $x = 0.6$ to
\unit[1.24]{mm} in figure~\ref{fig:correlation}(a).  The bubble size oscillates around the average value predicted by the estimate \eqref{eq:Rb}, $L_{\text{acc}}\approx\unit[9.5]{\micro m}$.  Electron
self-injection into the oscillating bubble leads to the formation of a quasi-monoenergetic component in the energy spectrum.  At the end of Stage I, at $z\approx\unit[1.24]{mm}$, the bubble contracts
to the same size in both runs, truncating the tail of injected bunch and expelling electrons injected between $x = 0.825$ and $x\approx\unit[0.95]{mm}$.  These electrons do not reach dephasing and
thus are missing in figures \ref{fig:correlation}(b) and \ref{fig:correlation}(c).  Electrons injected between $x = 0.65$ and \unit[0.825]{mm}, remain in the bubble and are further accelerated.  This
well-separated group of particles is clearly seen in figure \ref{fig:correlation}(b).  In both the \vorpalpd and \calder simulations, these electrons reach dephasing first, preserving low energy
spread, and are accelerated to the highest energy, $E\approx\unit[500]{MeV}$.  The bubble expands more rapidly and stabilises sooner in the \vorpalpd simulation, causing stronger reduction of the
phase velocity in the subsequent buckets (second and third).  Hence, in contrast to the \calder run, \vorpalpd gives a noticeable amount of charge trapped and preaccelerated in these buckets.  These
electrons, indicated by the red ellipse in figure \ref{fig:correlation}(b), are swallowed by the expanding first bucket during Stage II and are further accelerated, contributing to the dark current.  This
contribution, however, appears to be fairly minimal in comparison to the amount of continuously injected charge during Stage II.

The leading edge of the laser pulse constantly experiences a negative gradient of the nonlinear index of refraction.  As a result, by the end of Stage I, it accumulates considerable redshift.  During
Stage II, plasma-induced group velocity dispersion slows the red-shifted spectral components relative to the unshifted components, leading to the front etching and pulse self-compressing into a
relativistically intense, few-cycle long optical piston \citep{Tsung:PulseCompression,Lontano:Dynamics,Faure:PulseCompression,Kalmykov:SelfInjectionPhysPlasmas}.  As the pulse transforms into a
piston, the bubble constantly elongates, resulting in copious trapping and creating a poorly collimated, polychromatic tail, clearly seen in figure \ref{fig:correlation}(b).  At the dephasing point,
$x_{\text{deph}}\approx\unit[2.4]{mm}$, the bubble size becomes nearly twice the estimate $L_{\text{acc}}\approx\unit[9.5]{\micro m}$ based on the scaling law \eqref{eq:Rb}.  Even though figure
\ref{fig:correlation}(a) shows a larger bubble expansion in the \vorpalpd run, the sections of collection phase space corresponding to Stage II look nearly identical for both codes in figure
\ref{fig:correlation}(b).

The collection volume depicted in figure \ref{fig:correlation}(c) indicates that the electrons are collected from a conical shell with a radius slightly smaller than the bubble radius.  This structure
of the collection volume indicates that the vast majority of trapped and accelerated electrons have impact parameters of sheath electrons
\citep{Tsung:Simulation,Wu:InjectionFromSheath,Pukhov:InjectionFromSheath,Kalmykov:NumericalModelling,Kalmykov:SelfInjectionPhysPlasmas}.  Collection volumes in the \vorpalpd and \calder runs are
almost identical during Stage I, whereas the radius of the cone is larger for \vorpalpd during Stage II, on account of the greater expansion due to pulse diffraction.

\begin{figure}
	\begin{center}
		\includegraphics{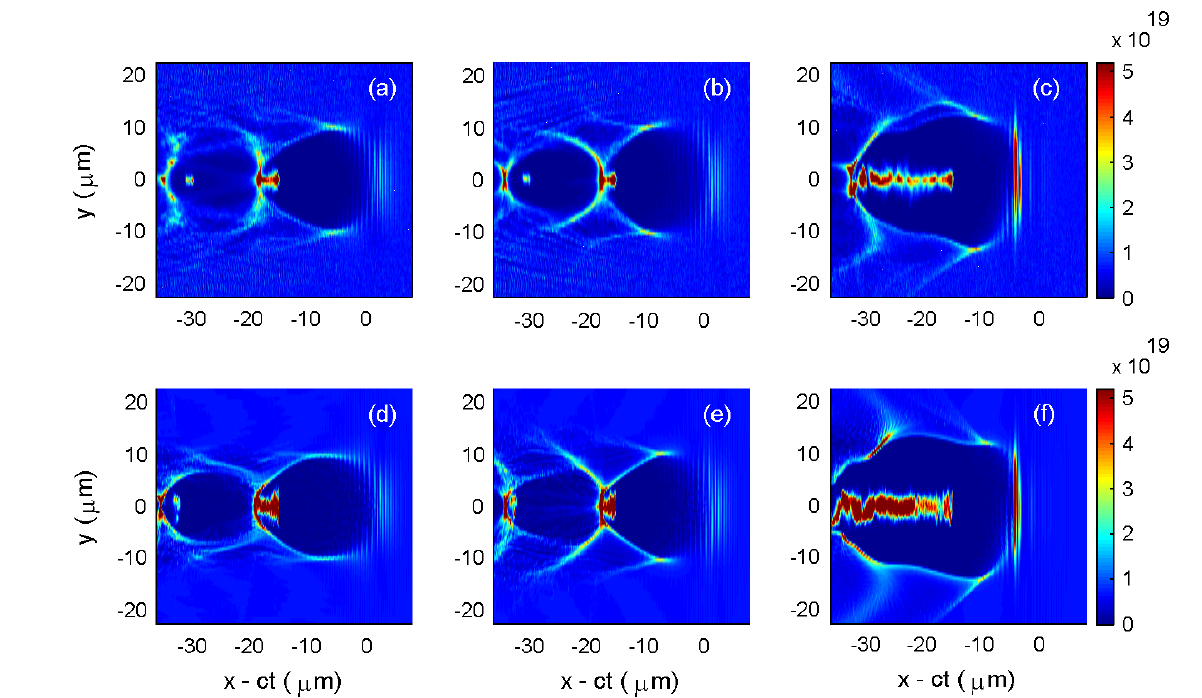} 
		\caption{Electron density (in $\unit{cm}^{-3}$) in the plane of laser polarisation in \calder (top row) and \vorpalpd simulations (bottom).  Panels (a) and (d) show the fully expanded bubble in the
middle of Stage I,
		(b) and (e) the fully contracted bubble at the end of Stage I, and (c) and (f) the bubble in the vicinity of electron dephasing point at the end of Stage II. $x = ct$ is the trajectory of the laser
		pulse maximum in vacuum; (a) and (d) correspond to the distance $x = ct \approx\unit[930]{\micro m}$ from the plasma edge, (b) and (e) to $x = ct\approx\unit[1210]{\micro m}$, and (c) and (f) to $x
		= ct\approx\unit[2364]{\micro m}$.  Before the dephasing point, the bubble, elongated and deformed due to the laser pulse self-compression, traps considerable charge.  Beam loading, however, is yet
		unable to terminate self-injection (cf.\ panels (c) and (f)).}
		\label{fig:density}
	\end{center}
\end{figure}

\begin{figure}
	\begin{center}
		\includegraphics{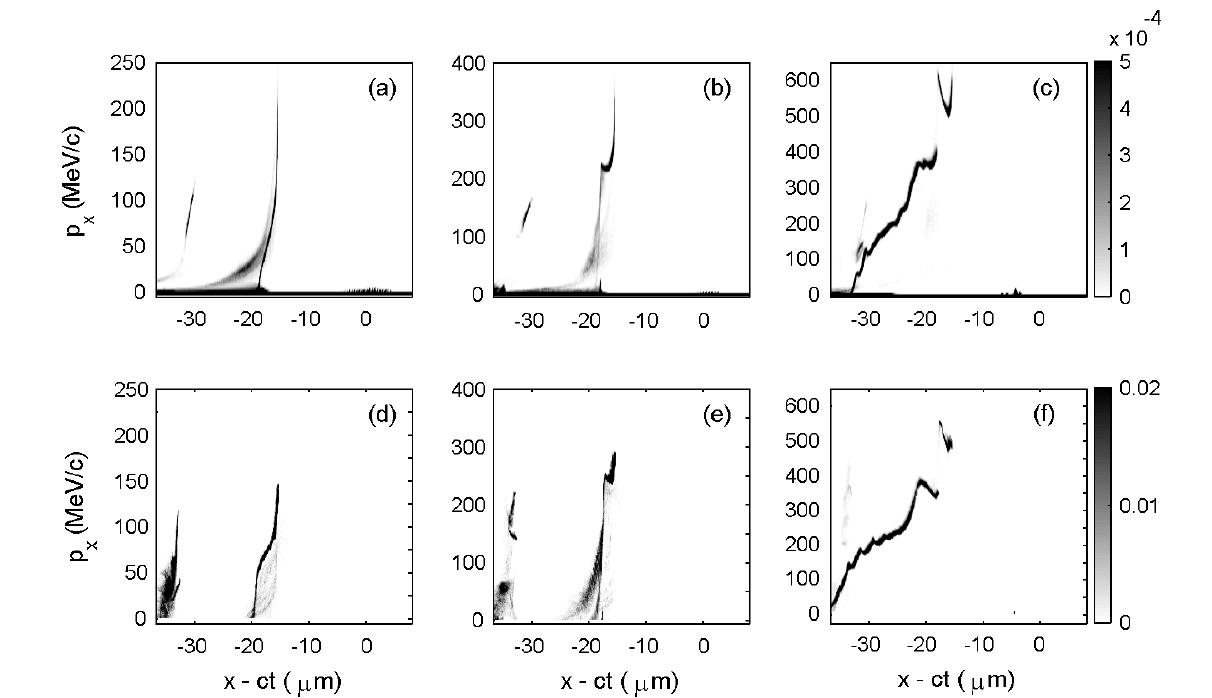} 
		\caption{Electron density (arbitrary units) in longitudinal phase space in \calder (top row) and \vorpalpd simulations (bottom).  Each panel corresponds to the same panel of
		figure~\ref{fig:density}.  Full expansion of the bubble saturates injection and initiates phase space rotation (panels (a) and (d)).  Contraction of the bubble terminates injection, clipping the
		rear of injected bunch, eliminating low-energy tail.  Phase space rotation makes the bunch quasi-monoenergetic (panels (b) and (e)).  Elongation and deformation of the bubble due to the laser pulse
		self-compression causes continuous injection, producing an electron beam with a continuous spectrum of longitudinal momenta (panels (c) and (f)).}
		\label{fig:pspace}
	\end{center}
\end{figure}

\begin{figure}
	\begin{center}
		\includegraphics{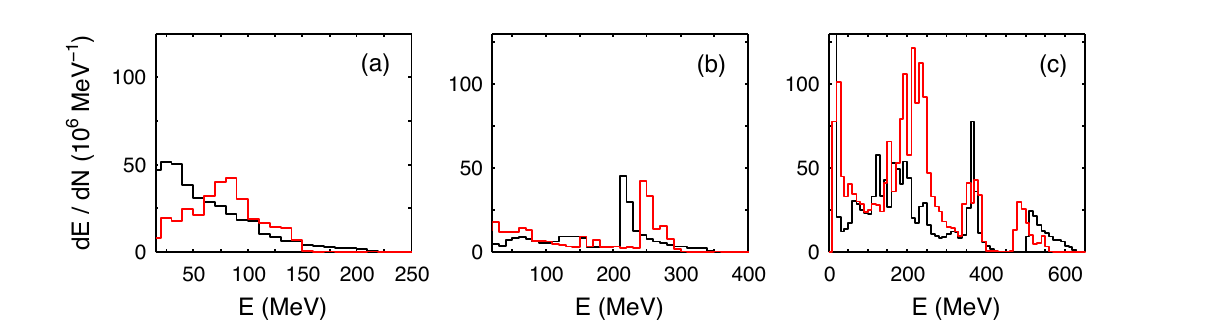} 
		\caption{Electron energy spectra in \calder (black) and \vorpalpd simulations (red/grey).  Panels (a), (b), and (c) correspond to the phase space snapshots (a) and (d), (b) and (e), and (c) and (f)
		of figure~\ref{fig:pspace}, respectively.  (a) At the point of full expansion, the electron energy spectrum is broad.  (b) Full contraction of the bubble suppresses the low-energy tail and reduces
		the energy spread.  Electrons from the second bucket contribute to the background, seen in the diffuse peaks around \unit[150]{MeV}.  (c) Continuous injection caused by the bubble expansion and
		deformation produces a massive polychromatic tail.  The leading bunch, at $E\approx\unit[500]{MeV}$, reaches dephasing, but is still distinct from the tail.}
		\label{fig:energy}
	\end{center}
\end{figure}

Snapshots of electron density, longitudinal phase space, and energy spectra at the points of maximal expansion and contraction of the bubble are presented in figures \ref{fig:density},
\ref{fig:pspace},
and \ref{fig:energy}.  Data for panels (a), (b), and (c) in these figures are from the \calder simulation, and for panels (d), (e), and (f) from the \vorpalpd simulation.

The fully expanded bubble in the middle of Stage I is shown in figures \ref{fig:density}(a) and \ref{fig:density}(d).  As soon as the bubble fully expands, injection terminates.  Uninterrupted
injection of
sheath electrons before this point produces a large spread of longitudinal momentum and energy, shown in figures \ref{fig:pspace}(a), \ref{fig:pspace}(d), and \ref{fig:energy}(a).

The slight contraction of the bubble between $x = 0.95$ and \unit[1.24]{mm} truncates the bunch.  Electrons injected at the very end of the expansion interval are expelled, while particles remaining
in the
bucket are further accelerated.  The transverse self-fields of the bunch are unable to prevent the bucket contraction.  Snapshots of the contracted bubble are presented in figures
\ref{fig:density}(b) and
\ref{fig:density}(e).  During the contraction interval, the tail of electron bunch, exposed to the highest accelerating gradient, equalises in energy with earlier injected electrons, thus producing a
characteristic `U' shape in the longitudinal phase space.  This feature (also observed in the similar situation by \citep{Lu:Generating}) is clearly seen in figures \ref{fig:pspace}(b) and
\ref{fig:pspace}(e).  As a result of this evolution, quasi-monoenergetic bunches form in both \vorpalpd and \calder simulations at the end of Stage~I. These quasi-monoenergetic spikes with $<10\%$
energy
spread can be seen in figure \ref{fig:energy}(b).  In addition to the quasi-monoenergetic spikes, these energy spectra also reveal diffuse features near \unit[150]{MeV}, corresponding to the electrons
trapped in the second bucket; these particles can be seen in the snapshots of electron density shown in figures \ref{fig:density}(b) and \ref{fig:density}(e).  These electrons, however, never equalise in energy with
the
leading high-energy bunch.

\begin{table}
	\begin{center}
		\def~{\hphantom{0}}
		\begin{tabular}{lcccccc}
			& $Q_{\text{mono}}$ & $E_{\text{mono}}$ & $\Delta E_{\text{mono}}$  & $\varepsilon_{N,y}$ & $\varepsilon_{N,z}$ \\[3pt]
			\calder  & 214 & 215 & 20  &  6.87  & 7.08\\
			\vorpalpd   & 193 & 245 & 20 & 10.7 &  6.08 \\
		\end{tabular}
		\caption{Parameters of the quasi-monoenergetic bunch ($E > \unit[200]{MeV}$) at the end of Stage I (cf.\ the spectra in figure \ref{fig:energy}(b)).  $Q_{\text{mono}}$ is the charge in pC;
		$E_{\text{mono}}$ is the energy corresponding to the spectral peak (in MeV); $\Delta E_{\text{mono}}$ is the absolute energy spread (FWHM) in MeV; $\varepsilon_{N,y}$ and $\varepsilon_{N,z}$ are
		the normalised transverse emittance (in mm mrad) in and out of the laser polarisation plane, respectively.}
		\label{tab:parameters}
	\end{center}
\end{table}

Parameters of the bunches, summarised in table \ref{tab:parameters}, appear to be very similar.  Normalised transverse emittances presented in this table are calculated according to the usual
definition
$\varepsilon_{N,i}=(m_ec)^{-1}[ (\langle p_i^2\rangle - \langle p_i\rangle^2)(\langle r_i^2\rangle - \langle r_i\rangle^2) - (\langle p_i r_i \rangle - \langle r_i \rangle\langle p_i\rangle)^2
]^{1/2}$,
where $i=y$ and $z$ correspond to the emittance in and out of polarisation plane.  The beam asymmetry is more pronounced in the \vorpalpd simulation, presumably on account of the inclusion of the
complete
electromagnetic field, in contrast to just two poloidal modes in \calder.

Agreement between the two codes worsens during Stage II. As has already been noted, the bubble expansion is larger in the \vorpalpd simulation.  As a result, the amount of continuously injected charge
at the dephasing point (2.5 nC) is about 60\% higher and the divergence of the continuously injected beam (80 mrad) is about twice that in the \calder simulation.  The difference in charge can
be easily inferred from figure \ref{fig:energy}(c).  On the other hand, parameters of the leading bunches are in reasonable agreement, with the central energy $\unit[485\pm20]{MeV}$ in \vorpalpd
against $\unit[515\pm25]{MeV}$ in \calder run.  In both simulations, the emittance of the quasi-monoenergic component increases by $\about 30\%$ over its value at the end of Stage I. The lower
energy of the leading bunch in the \vorpalpd run can be explained by its earlier dephasing due to more rapid expansion of the bubble.

Both codes agree that the bubble not only elongates during Stage II, but becomes more and more asymmetric in the laser polarisation plane.  The ``pennant-like'' bubble shape is
responsible for massive off-axis injection, leading to the noticeable beam centroid oscillations in the laser polarisation plane seen in figures \ref{fig:density}(c) and \ref{fig:density}(f).  Such
phenomenon has been observed in similar situations by others \citep{Glinec:BetatronOffAxis}.  This violation of symmetry is a manifestation of carrier-envelope phase effects in the interaction of a
relativistically intense, linearly polarised, few-cycle piston with the plasma \citep{Nerush:CarrierEnvelope}.  Conversely, in the plane orthogonal to the laser polarisation, both the bucket and the
beam remain perfectly symmetric (not shown).  Surprisingly, the two poloidal radiation modes of \calder still capture the field evolution well.  Inclusion of higher order modes should improve the
situation.  On the other hand, figures \ref{fig:density}(c) and \ref{fig:density}(f) indicate that electromagnetic solvers of both codes agree on the group velocity of the laser pulse even in the
situation where the pulse shrinks down to less than two cycles and remains strongly relativistic.  This means that (a) poloidal mode decomposition does not damage dispersion in the axial direction,
and (b) the coarse grid and dispersion properties of \vorpalpd are sufficient to describe well the extreme case of pulse spectral broadening to $\Delta\omega\sim\omega_0$ and compression to nearly a
single cycle.

Examination of the bubble evolution and collection volumes (cf.\ figure \ref{fig:correlation}), together with individual snapshots of electron density in coordinate and longitudinal phase space,
indicate that, in spite of the great difference in the algorithms, \vorpalpd and \calder reproduce the same correlation between the evolution of the bubble and the self-injection of sheath electrons,
and agree quantitatively on the parameters of quasi-monoenergetic beams produced by the oscillating bubble.  Self-injection begins, terminates, and resumes at exactly the same positions along the
propagation axis in both runs, and electrons are collected from the same plasma volume.  Despite differences in minor details, both codes consistently reproduce physical details of the self-injection
process over the entire dephasing length.  This level of agreement between very different numerical models indicates that the results are largely free of numerical artefacts.  Importantly, the
discrepancies emerge when the interaction develops noticeable non-cylindrically symmetric features, and hence the reduced field description of \calder loses precision.  We believe that the agreement
between the models may be improved in a straightforward fashion (viz.\ using a larger number of poloidal modes) without significantly reducing computational efficiency.

\subsection{Effects of numerical dispersion control\label{sec:dispersionComparison}}

As described above, simulating upcoming experiments will require economising on computational cost as much as possible without sacrificing physical accuracy.  One means of reducing longitudinal
resolution
requirements, and hence allowing a larger time step, is to minimise numerical dispersion through a modified electromagnetic update.  Here we show how numerical dispersion quantitatively affects the
injected
electron bunch.

\begin{figure}
	\begin{center}
		\resizebox{3in}{!}{\includegraphics{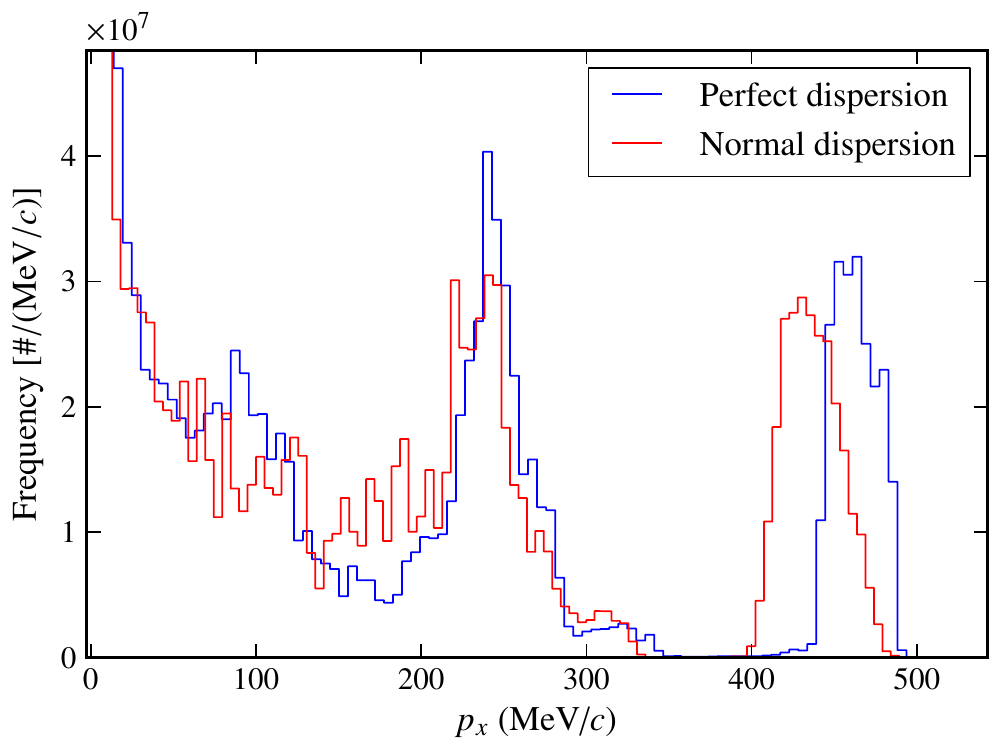}}\\
		\resizebox{\textwidth}{!}{\includegraphics{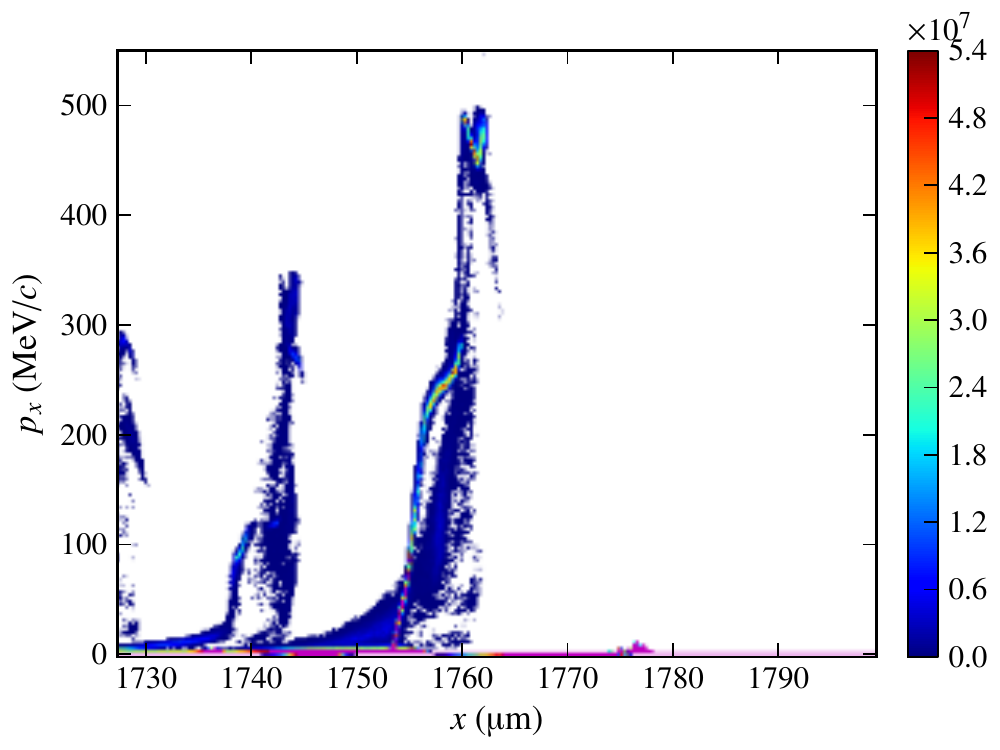}\ \includegraphics{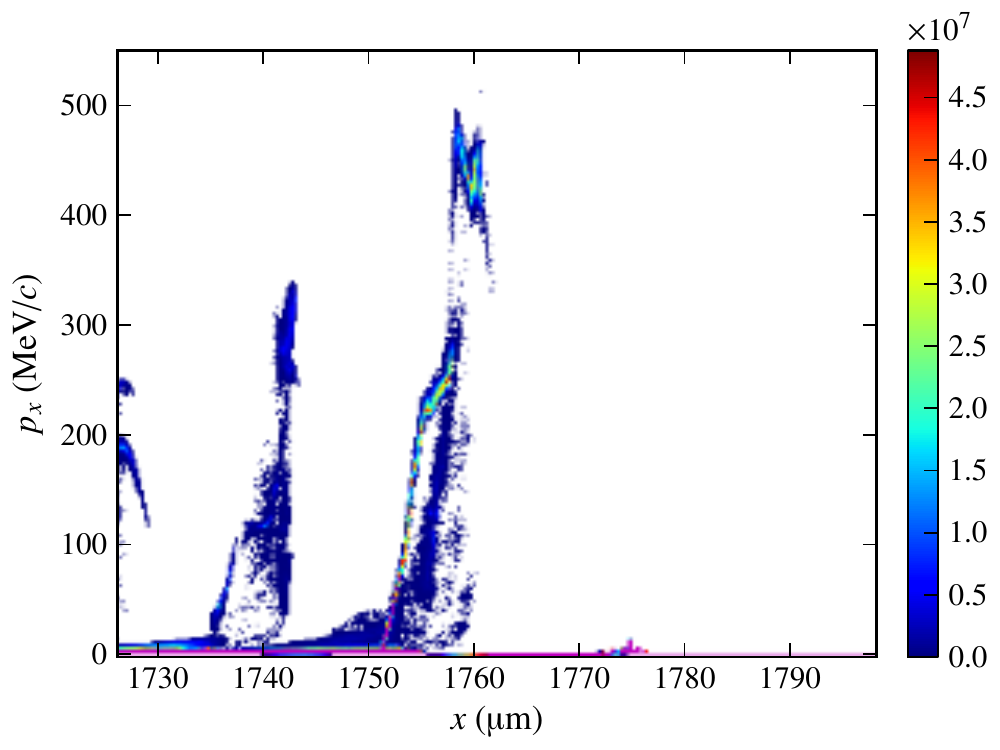}}
		\caption{Top: Electron longitudinal momentum spectra after \unit[1.8]{mm} propagation for perfect and normal dispersion.  Bottom: Longitudinal phase space for perfect dispersion (left)
		and normal dispersion (right).}
		\label{fig:dispersionComparison}
	\end{center}
\end{figure}

The immediate effect of numerical dispersion is an unphysically low group velocity of the laser pulse.  While this effect is difficult to observe directly in the laser pulse because of more
significant
changes in the pulse shape, it can be seen in the electron phase space, which is of experimental importance.  We examine the initially-injected electrons at the point where they have rotated in phase
space
such that the beam has achieved minimal energy spread.  The minimal energy spread condition is characterised by the phase space of the bunch being roughly longitudinally symmetric and in the shape of
a
`U'.  We find from the perfect dispersion simulation that this occurs after the laser has propagated approximately \unit[1.8]{mm} into the plasma.  We show longitudinal momentum spectra and phase
space at
this point for both perfect dispersion and normal dispersion in figure~\ref{fig:dispersionComparison}.
We find that with the normal dispersion algorithm, the beam achieves lower energy and exhibits higher energy spread than with the perfect dispersion algorithm.  We also find that phase space rotation has
occurred more quickly.

\begin{figure}
	\begin{center}
		\resizebox{\textwidth}{!}{\includegraphics{pdPhaseSpace45.pdf}\ \includegraphics{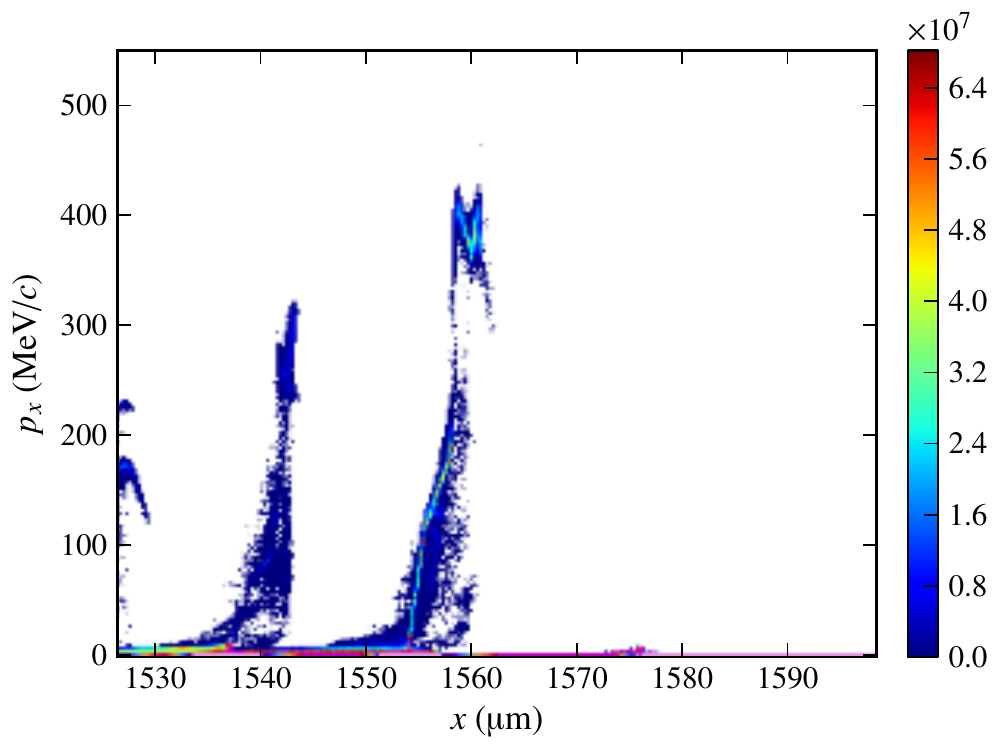}}
		\caption{Longitudinal phase space for perfect dispersion (left) and normal dispersion (right) at the minimal energy spread point for both algorithms.  For perfect dispersion, this is after
		\unit[1.8]{mm} of propagation, and for normal dispersion after \unit[1.6]{mm}.}
		\label{fig:minEnergySpread}
	\end{center}
\end{figure}

\begin{figure}
	\begin{center}
		\resizebox{\textwidth}{!}{\includegraphics{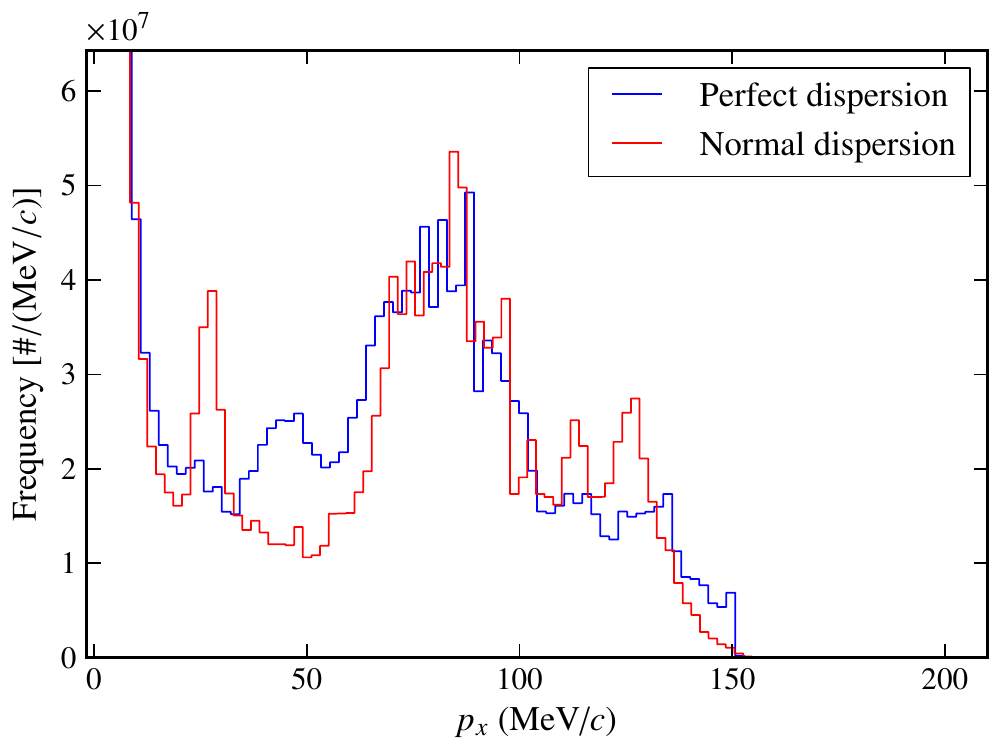}\ \includegraphics{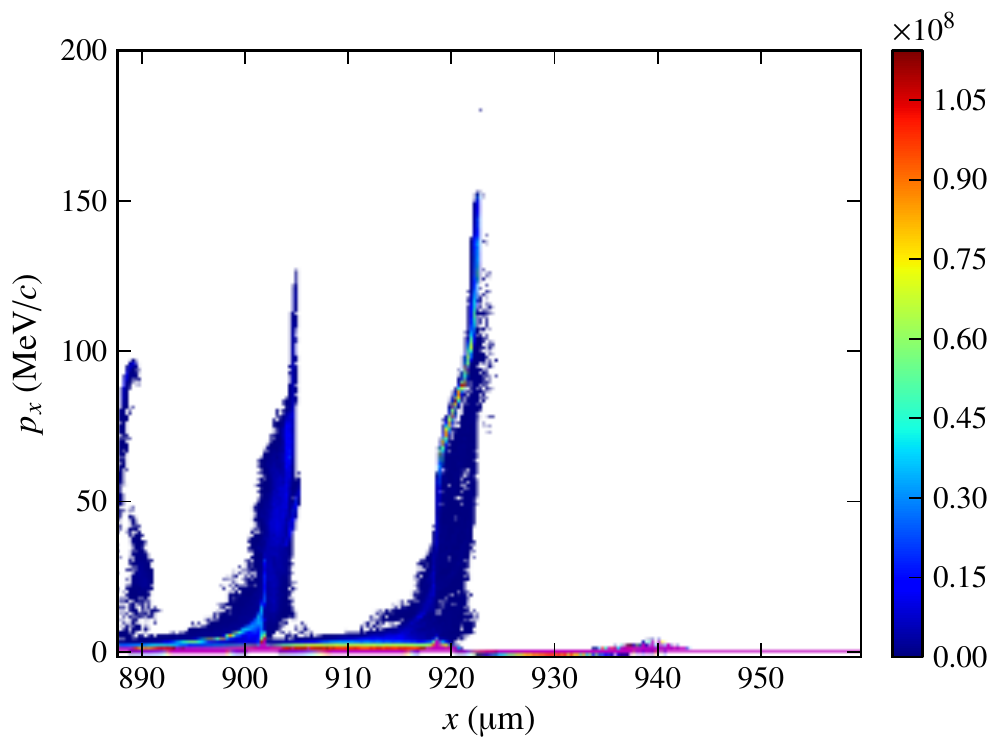}\ \includegraphics{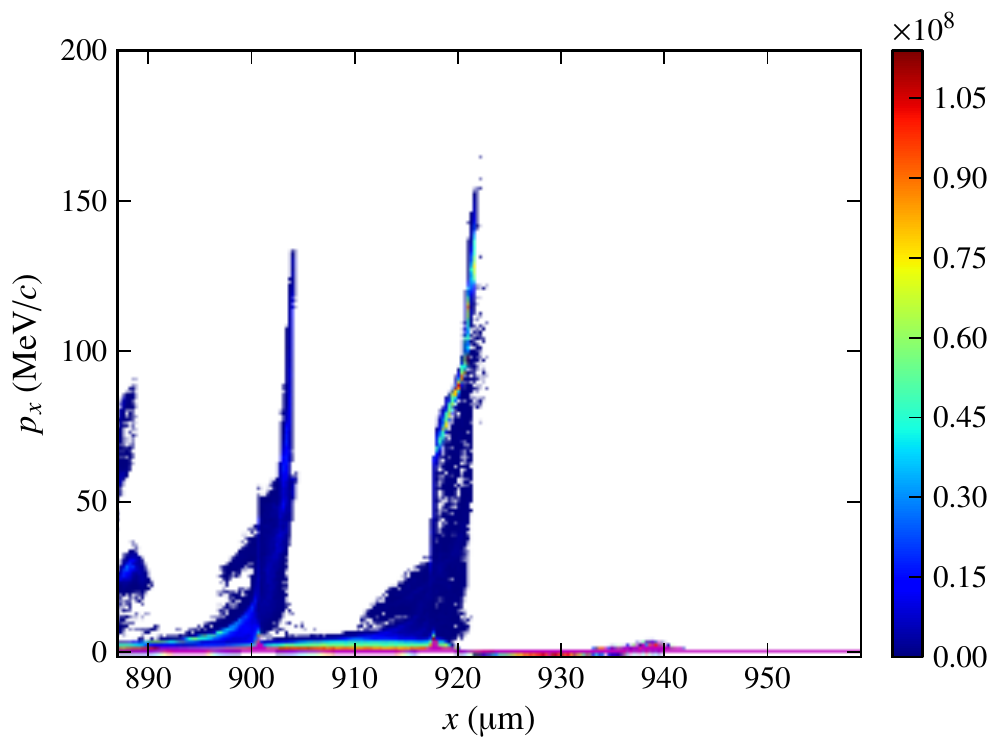}}\\
		\resizebox{\textwidth}{!}{\includegraphics{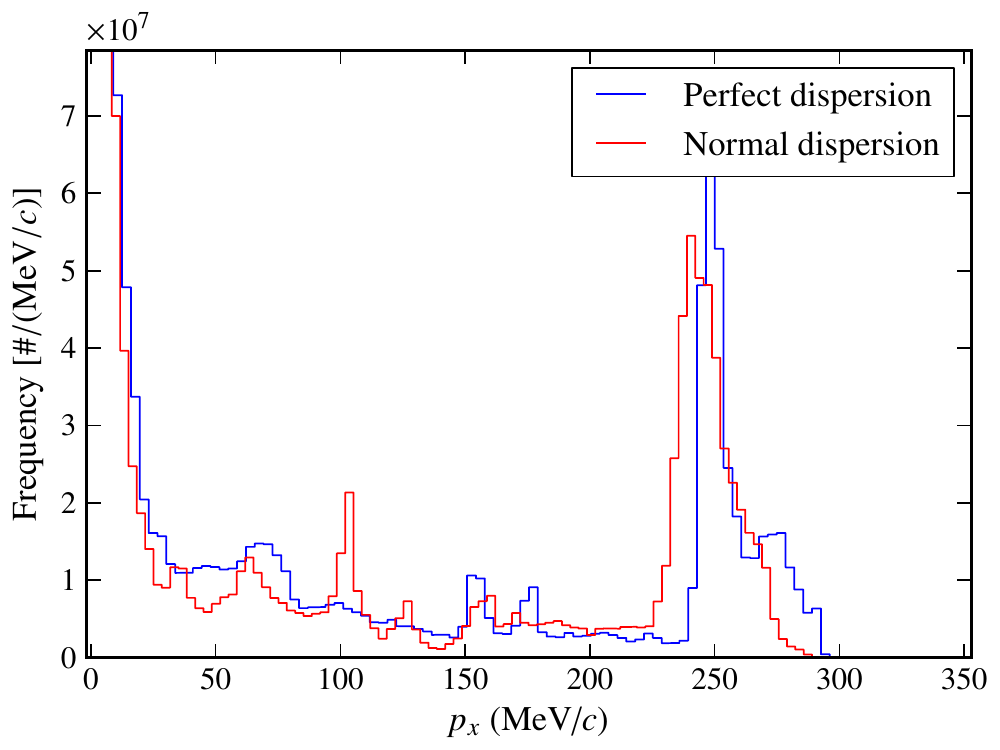}\ \includegraphics{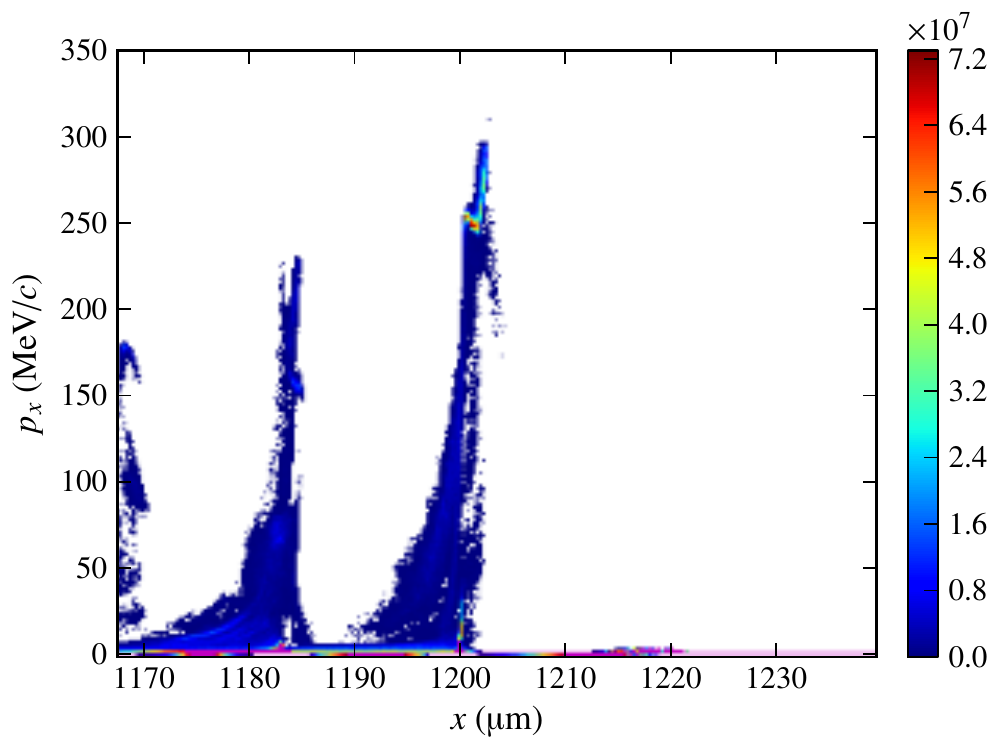}\ \includegraphics{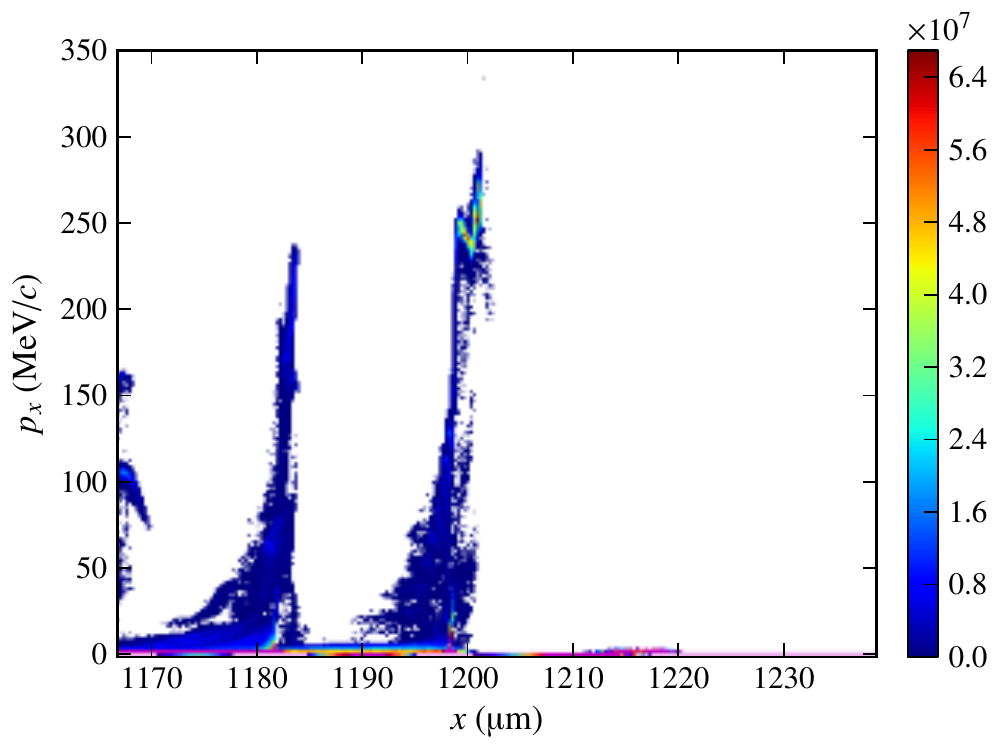}}\\
		\resizebox{\textwidth}{!}{\includegraphics{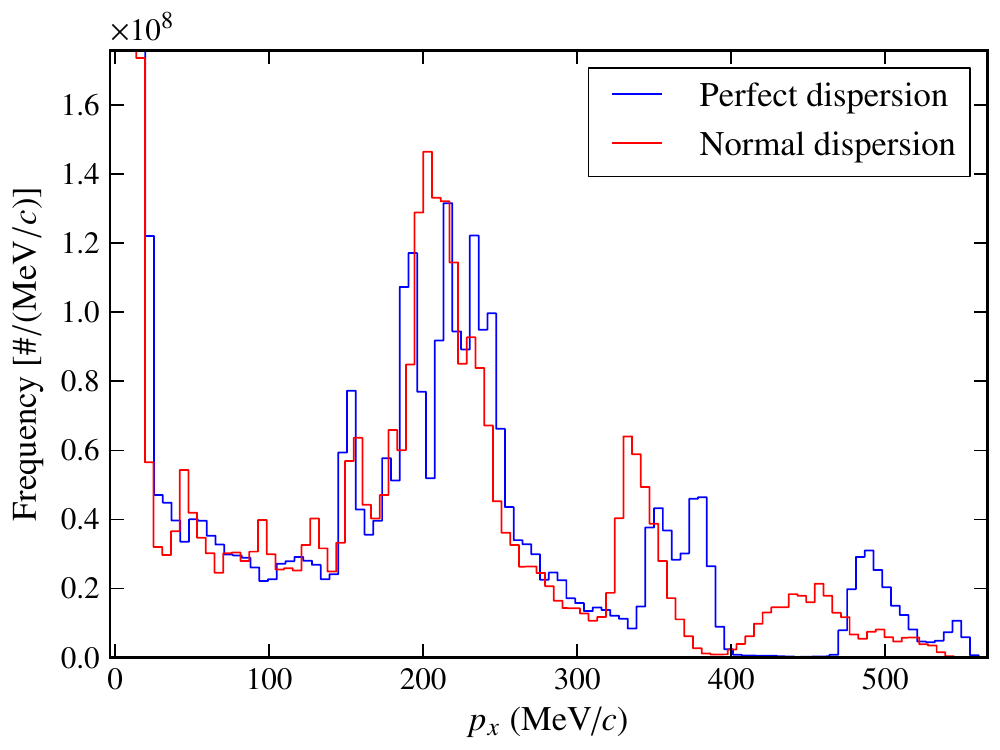}\ \includegraphics{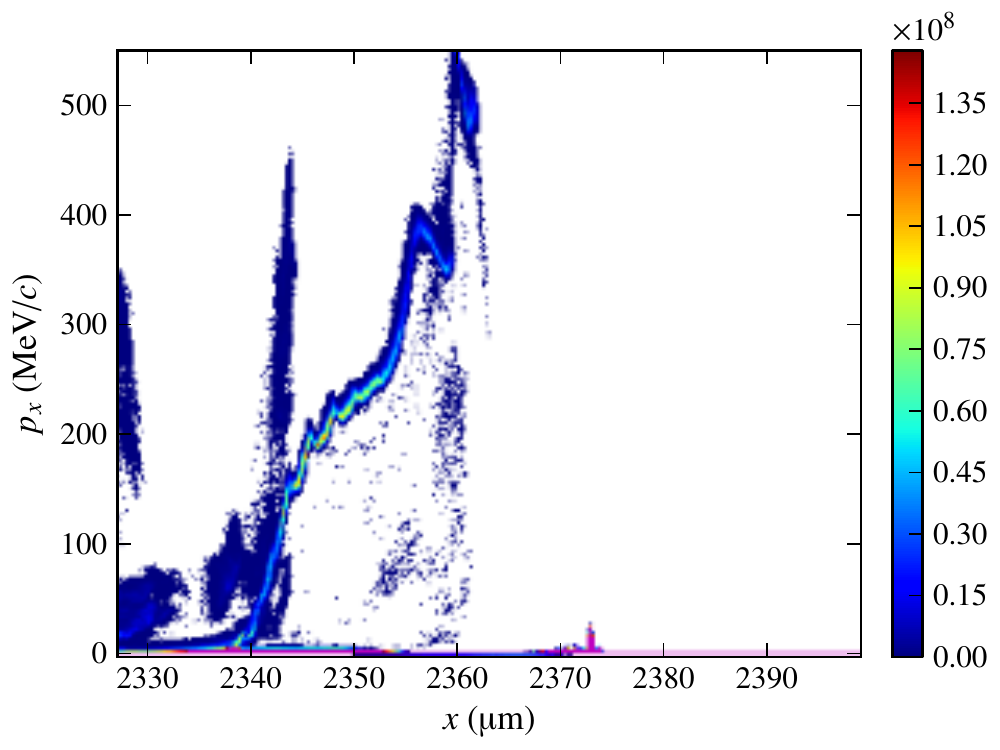}\ \includegraphics{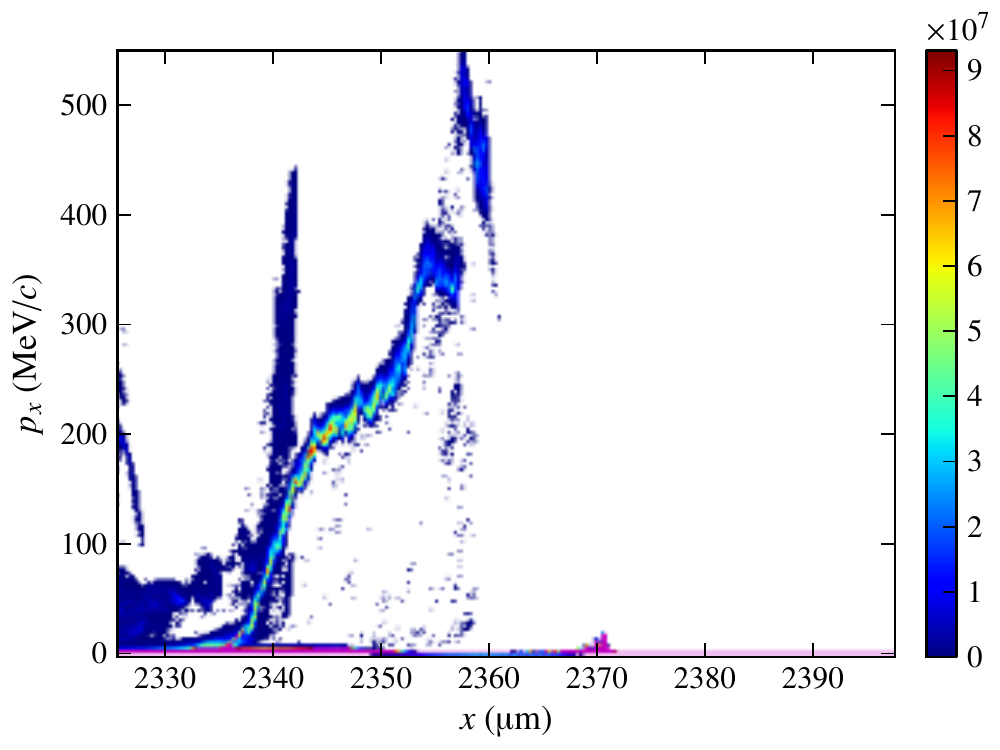}}\\
		\caption{Comparisons of longitudinal momentum and phase space.  Momentum spectra for perfect and normal dispersion (left), the perfect-dispersion phase space (center),
		and the normal-dispersion phase space (right).  The top row shows the electron distribution at $\unit[960]{\micro m}$ of propagation, the middle row at \unit[1.24]{mm}, and the bottom row
		at \unit[2.4]{mm}.}
		\label{fig:CalderPointComparisons}
	\end{center}
\end{figure}

We also compare the two dispersion algorithms at points of minimal energy spread.  Without dispersion control, the more rapid phase space rotation causes the minimal energy spread to occur after just
\unit[1.6]{mm} of propagation rather than \unit[1.8]{mm}.  We show the two phase space plots in figure~\ref{fig:minEnergySpread}.  This comparison is relevant since for applications, as one would
want to
design the system such that the injected beam exits the plasma at this point of minimum energy spread \citep{Hafz:2011aa}.  It is clear from these plots that with the normal dispersion algorithm, the
beam has
reached lower mean energy (\unit[390]{MeV}) at the minimum energy spread point than with perfect dispersion, where the beam has mean energy of \unit[460]{MeV}.  In addition, the normal dispersion case
exhibits slightly higher energy spread and total charge in the bunch.

We also compare the longitudinal momentum spectra and phase space at the points compared with \calder simulations in the previous section, namely $\unit[960]{\micro m}$, \unit[1.24]{mm}, and
\unit[2.4]{mm}.  These comparisons are shown in figure~\ref{fig:CalderPointComparisons}.  We can see, especially in the later two plots, that the injected bunch in the normal dispersion simulation
shows both
lower mean energy and greater phase space rotation than in the perfect dispersion run, which agrees better with the \calder simulations as seen in the previous section.

While these discrepancies are small, they are noticeable and consistent with numerical group velocity error.  As LPA system designs are refined, and diagnostics and control over the laser pulse and
plasma
improve, it will be important to control numerical effects on this level to optimise parameters through simulation.  The perfect dispersion algorithm allows us to do so while still using low
longitudinal
resolution for computational efficiency.

\section{Conclusions\label{sec:conclusions}}

In this paper we have demonstrated the utility of using computationally efficient, fully explicit 3D PIC codes to describe and explain the physical phenomena accompanying electron acceleration until
dephasing
in a self-guided LPA in the blowout regime.  Electron self-injection and its relation to nonlinear dynamical processes involving the laser pulse and bubble were explored.  Two approaches to reducing
the
computational cost of the simulations were considered.  

First, using the Cartesian code \vorpal\ with a newly developed perfect dispersion algorithm \citep{benc:PerfDispInPrep}, \vorpalpd, made it possible to use large grid spacings ($\about 15$ grid
points per
wavelength in the direction of propagation) and proportionally larger time steps.  This approach reproduces the correct group velocity of a broad-bandwidth laser pulse.  The red-shift,
self-compression, and depletion of the laser pulse were thus described correctly, with proper resolution of all important physical scales.

Second, the well-preserved axial symmetry of the problem allowed us to use a reduced geometry description, with poloidal-mode decomposition of currents and fields.  This approach was realised in the code
\calder \citep{Lifschitz:CalderCirc}.  By using only two modes, we approached the performance of a 2D code, at the same time preserving the correct cylindrical geometry of the interaction.  The high
computational efficiency of \calder allowed us to use a very high longitudinal resolution ($\about 50$ grid points per laser wavelength in the direction of propagation) and a large number of macroparticles
($\about 50$ per cylindrical cell), eliminating numerical dispersion and strongly reducing the sampling noise.  This high resolution simulation did not indicate any new physical effects relative to
the
\vorpalpd runs, and did not exhibit significant differences in the quantitative results.  Even with strong violation of cylindrical symmetry (such as near the dephasing limit, when the pulse was
transformed
into a two-cycle relativistic piston), the \calder results remained qualitatively correct.

Both codes described precisely the self-focusing of the laser pulse, the oscillations of its spot size, and related oscillations of the bubble; electron self-injection into the oscillating bubble and
formation of a quasi-monoenergetic bunch; laser pulse frequency broadening and self-compression into the relativistic piston; constant elongation of the bubble during the piston formation; and
uninterrupted electron injection eventually overloading the bubble.  The codes showed excellent agreement on the locations of initiation and extinction of injection, on the collection volume, and on
parameters of the quasi-monoenergetic component in the electron spectrum, indicating that the results are free of numerical artefacts.  It is especially interesting that the \calder simulation with
just
two poloidal modes did not lose accuracy and preserved the correct group velocity (agreeing with the \vorpalpd run) even when the laser pulse was compressed down two cycles.

We thus conclude that (1) using perfect dispersion, taking a coarser grid and larger time steps, and using higher-order splines for macroparticle shapes to suppress the sampling noise, or (2)
neglecting
high-order non-axisymmetric field and current components, thus reducing the dimensionality of problem are both effective and promising means to increase the computational efficiency without
sacrificing
fidelity.  Both of these methods are applicable to the design of upcoming experiments on GeV-scale acceleration of electrons with \unit[100]{TW}-scale lasers.

The work of B.~M.~C.\ and D.~L.~B.\ was partially supported by U.~S.\ DOE Contract No.\ DE-SC0006245 (SBIR).  The work of S.~Y.~K.\ and B.~A.~S.\ was partially supported by U.~S.\ DOE Contract No.\
DE-FG02-08ER55000 and by NSF Grant PHY-1104683.  X.~D.\ and A.~F.~L.\ thank Victor Malka for his support during the development of \calder.  A.~B.\ and E.~L.\ acknowledge the support of
LASERLAB-EUROPE/LAPTECH through EC FP7 contract no.\ 228334.  The \calder simulations in this work were performed using high-performance computing resources of GENCI-CCRT (grant 2010-x2010056304).
\vorpal\ simulations used resources of the National Energy Research Scientific Computing Center, which is supported by the Office of Science of the U.~S.\ DOE under Contract No.\ DE-AC02-05CH11231.

\bibliographystyle{jpp}
\bibliography{biblio}

\end{document}